\documentclass[12pt]{article}

\usepackage{graphicx}
\def\lsim{\mathrel{\rlap{
\lower4pt\hbox{\hskip-3pt$\sim$}}
    \raise1pt\hbox{$<$}}}     
\def\gsim{\mathrel{\rlap{
\lower4pt\hbox{\hskip-3pt$\sim$}}
    \raise1pt\hbox{$>$}}}     

\begin{document}
\begin{center}
{\Large \bf \boldmath Hydrodynamic View in the NICA Energy Range} 

\vspace*{6mm}
{V.~Toneev$^a$ and V.~Skokov$^a$ }\\      
{\small \it $^a$ Joint Institute for Nuclear Research, 141980 Dubna, Moscow Region, Russia }
\end{center}

\vspace*{6mm}

\begin{abstract}
Results for penetrating probes treated within a hybrid hydro-kinetic model are 
projected onto the energy range covered by the NICA and FAIR projects.  A new                                                                                                                                                                                                                                                   
source of dileptons emitted from a mixed quark-hadron phase, quark-hadron bremsstrahlung,
 is proposed. An estimate for the $\pi\pi \to\sigma \to \gamma\gamma$ process in 
 nuclear collisions is given. 
\end{abstract}

\vspace*{6mm}

D.I. Blokhintsev, whose centennial anniversary of the birthday this conference
is devoted to, has contributed to various fields of physics and its 
applications, particularly, to hydrodynamics. 
Sixteen years ago, at the dawn of hydrodynamics he made an important 
remark~\cite{Bl57} concerning possible violation of the uncertainty principle 
in initial conditions of the Landau hydrodynamic theory. From up-to-date view, 
this Blokhintsev's estimate looks slightly naive but in principle it is 
correct until now and should be taken into account in modern development 
of a relativistic hydrodynamic approach. Here we present 
a hybrid model which combines kinetic and hydrodynamic descriptions. To certain 
extent it can be considered  as a possible solution of the problem 
put by D.I.~Blokhintsev.
 
 In the hybrid model~\cite{ST06}, 
the initial stage of heavy ion collisions is treated  
kinetically within the transport Quark Gluon String Model (QGSM)~\cite{QGSM}  
whereas the subsequent stage is considered as an  isentropic expansion 
of a formed dense and hot system (fireball). The transition from one stage to another 
 is solved by considering the entropy evolution.
 
\begin{figure}[t]
\centerline{\includegraphics[width=6cm]{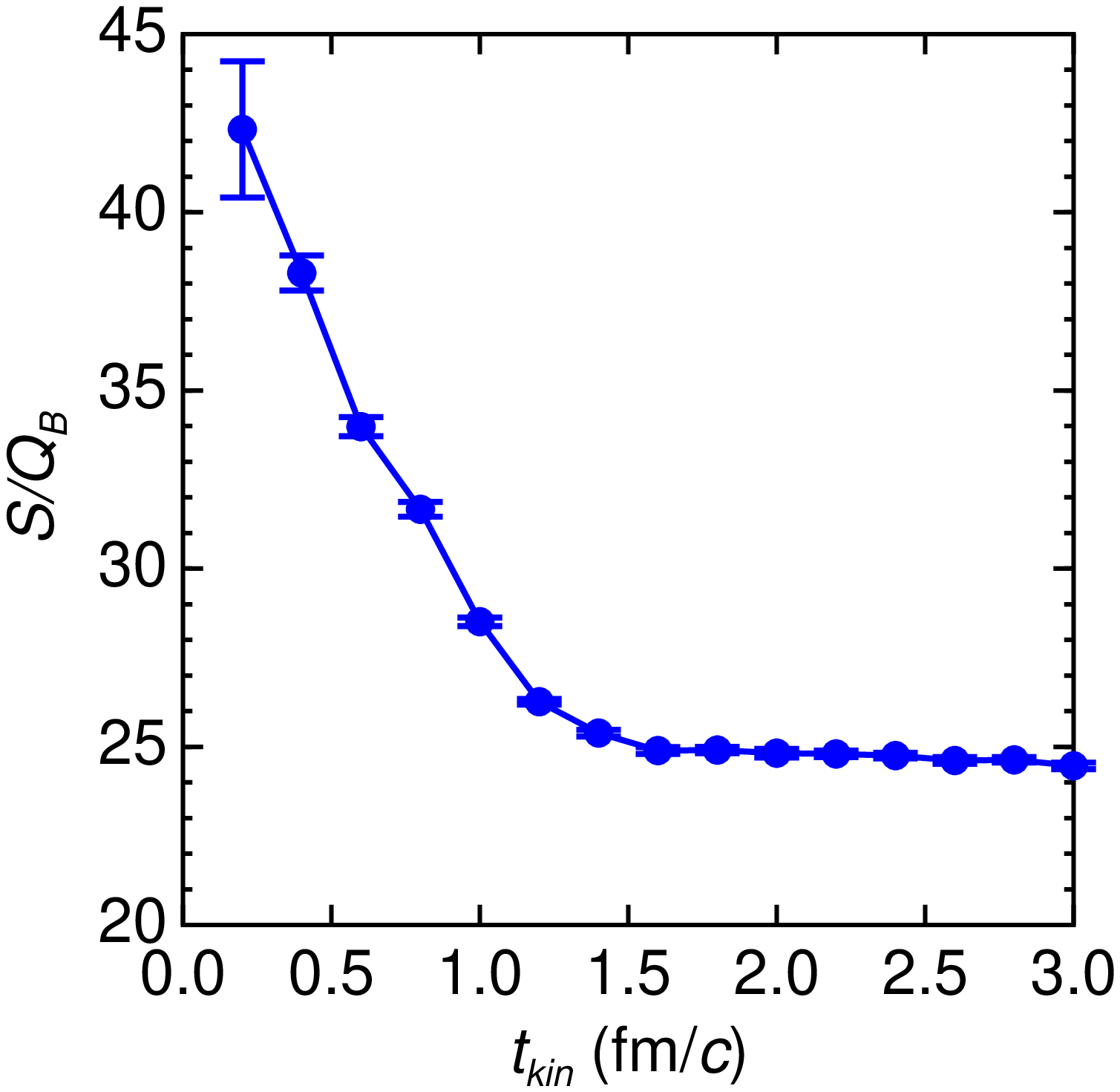}\hspace*{10mm}
\includegraphics[width=6cm]{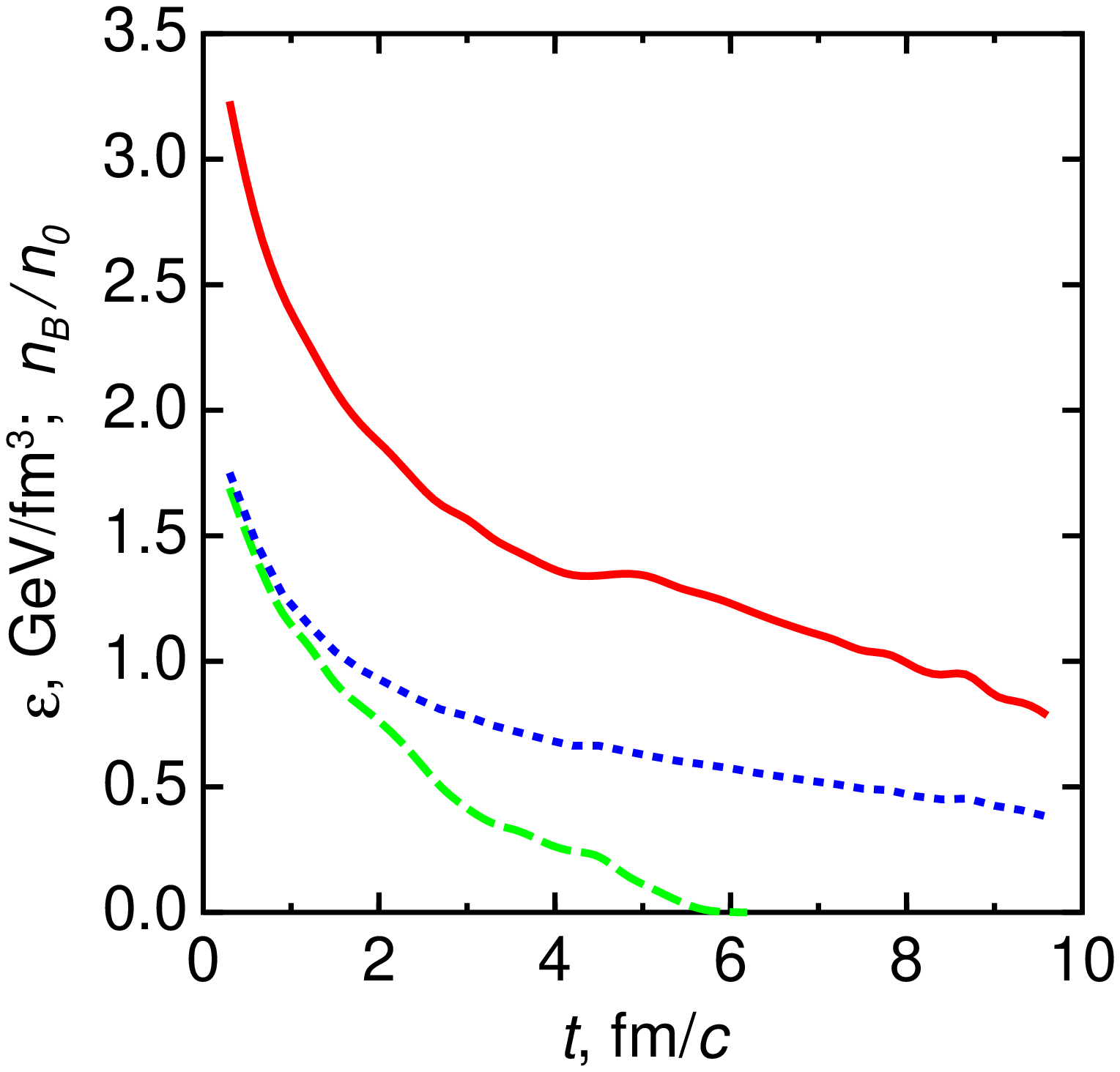}}
\caption{Temporal dependence of entropy $S$ per baryon charge $Q_B$ of participants 
(left panel) for a
semi-central In+In collision at $E_{lab}=$158 $A$GeV. In the right panel the average 
energy (solid line)   and baryon  (dashed) densities  of an expanding  
fireball formed in this collision. 
  Dotted line shows a contribution of quarks and gluons to the energy density.}
\label{sB}
\end{figure}

In Fig.\ref{sB}, the ratio between  entropy $S$ and baryon charge $Q_B$ of 
participants is shown for In+In collisions at the impact parameter $b=4$ fm and 
bombarding energy 158 $A$GeV.  Being calculated on a large 3D grid, this ratio is
less sensitive to particle fluctuation  as compared to the entropy
itself. Small values of the baryon charge $Q_B$ at the very beginning of collision
result in large values of the $S/Q_B$ ratio. It is clearly seen
that for $t_{kin}\gsim 1.3$ fm/$c$ this ratio is practically 
constant and this stage may be considered as isentropic expansion.

 To proceed from kinetics to hydrodynamics, we  evaluate
conserved components of the energy-momentum tensor $T_{00},
T_{01}, T_{02}, T_{03}$ and baryon density $n_B$ (the zero component of
the baryon current) within QGSM at the moment $t_{kin}=1.3$ fm/$c$
in every cell on the 3D grid. This state is treated as an initial
state for subsequent hydrodynamic evolution of a fireball. The
time dependence of average thermodynamic quantities is presented
in the left panel of Fig.\ref{sB}.

 The latter stage is evaluated within the relativistic 3D
hydrodynamics~\cite{ST06}. The key quantity is the equation of state.
In this work, the mixed phase Equation of State
(EoS) is applied~\cite{TFNFR04} which allows for coexistence of 
hadrons and quarks/gluons. This thermodynamically consistent
EoS uses the modified Zimanyi mean-field interaction for hadrons
and also includes interaction between hadron and quark-gluon
phases, which results in a crossover deconfinement phase
transition. In addition to~\cite{TFNFR04}, the hard thermal loop
term was self-consistently added to the interaction of quarks and
\begin{figure}[h]
\centerline{
\includegraphics[height=6cm,width=7cm]{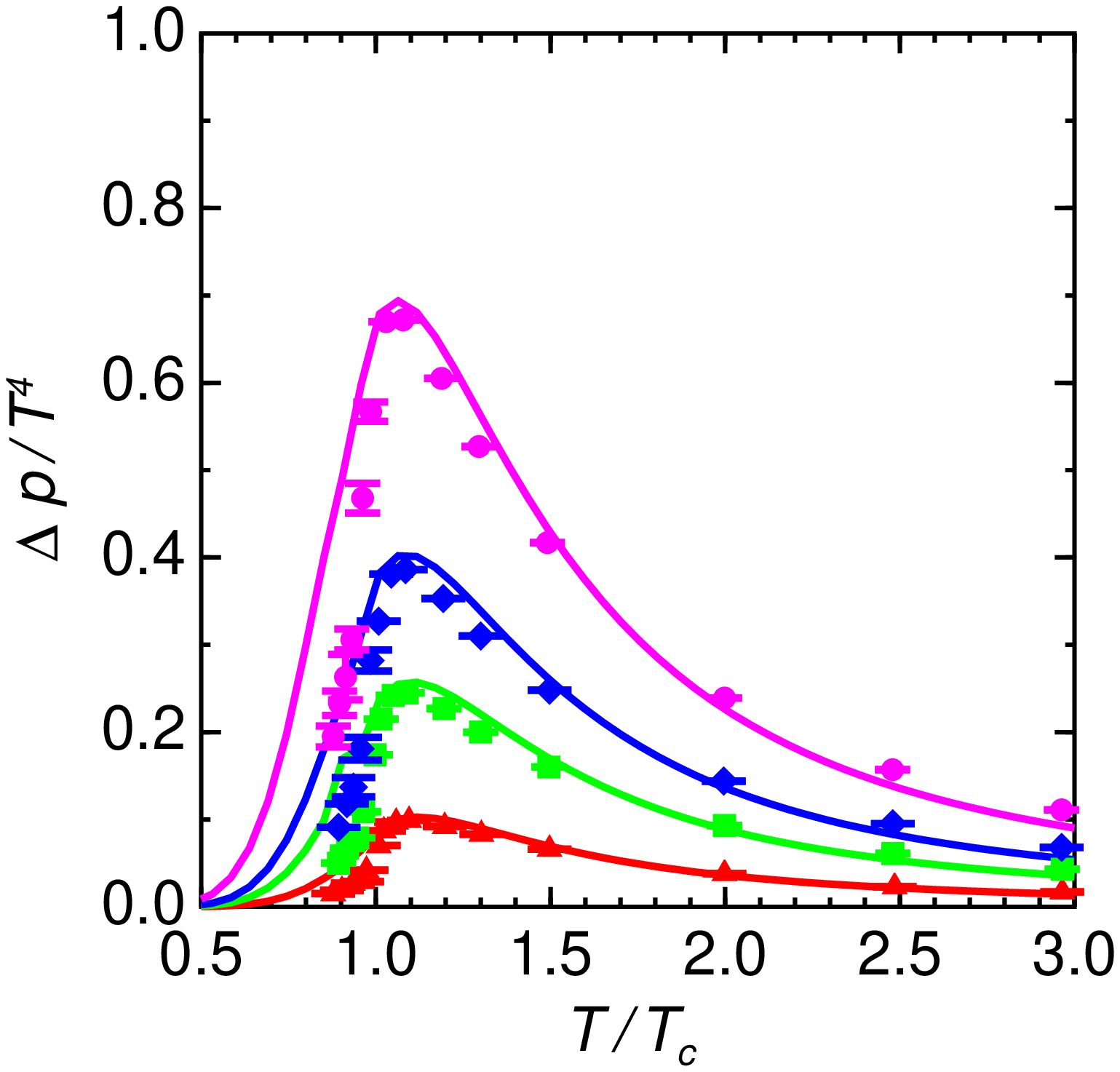}\hspace*{10mm}
\includegraphics[height=6cm,width=7cm]{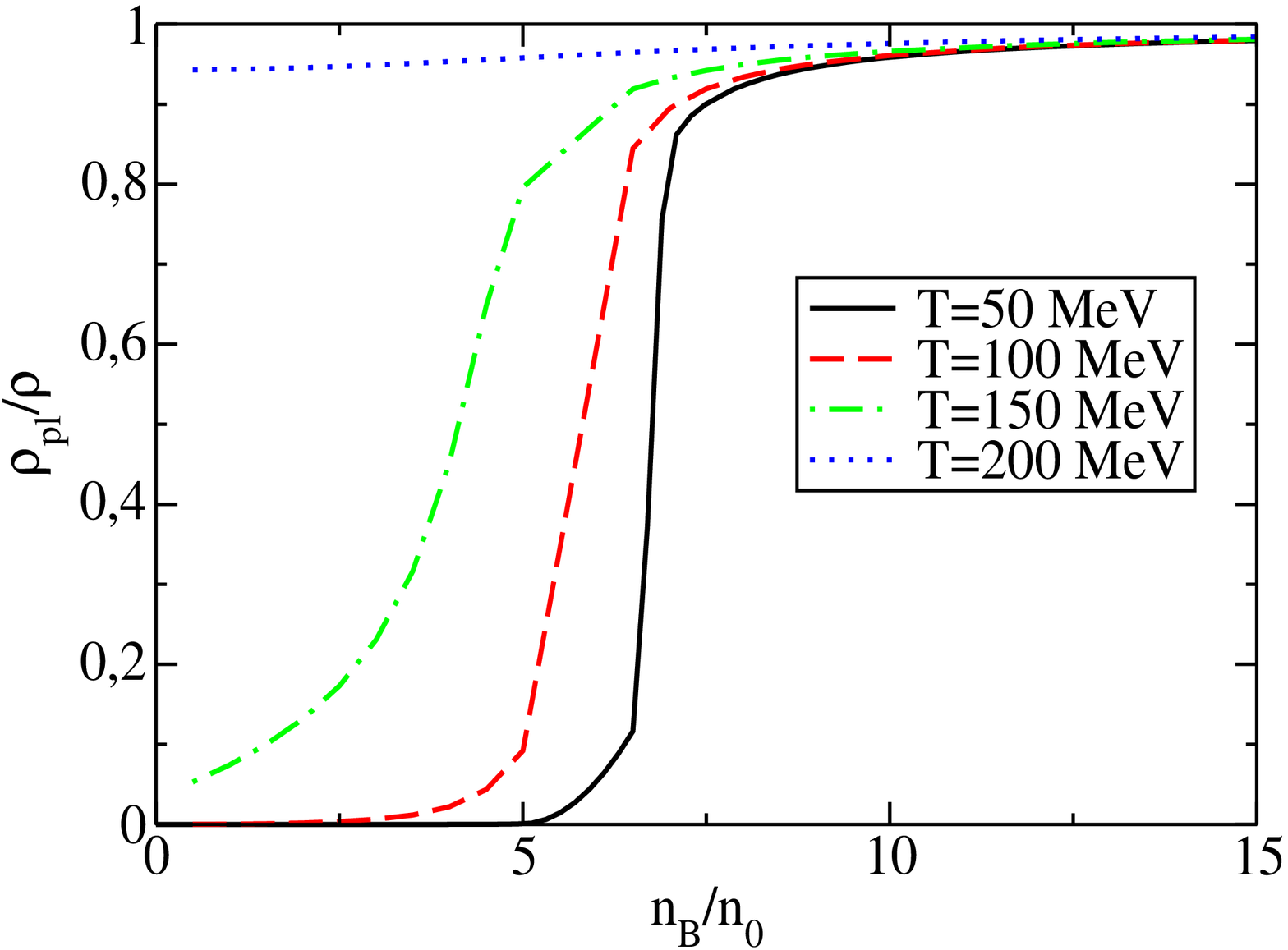} }
\caption{Temperature dependence of the reduced pressure (left panel)
 at the  baryon chemical potential $\mu_B=$
210, 330, 410 and 530 MeV (from
the bottom)  and fraction of unbound  quarks (right panel) 
within the mixed phase EoS. Points are lattice QCD data for
the 2+1 flavor system~\cite{Fodor03}. }
\label{lat}
\end{figure}
gluons to get the correct asymptotics at $T>>T_c$ and reasonable
agreement of the model results with lattice QCD calculations at
finite temperature $T$ and chemical potential
$\mu_B$~\cite{Fodor03}. This agreement is demonstrated in Fig.\ref{lat} where
the reduced pressure $\Delta p/T^4=(p(\mu_B)-p(\mu_B=0))/T^4$ is compared with recent 
lattice QCD data. 

The fraction of unbound quarks/gluons defined as $\rho_{pl}/\rho=(n_q 
+n_{\bar{q}}+n_g)/(n_q +n_{\bar{q}}+n_g+n_B+n_M)$ is presented for the mixed 
phase EoS in the right panel of Fig.\ref{lat}. It is seen that even 
at a moderate temperature $T\sim 50\div 100$ MeV the 
quark/gluon fraction sharply increases at the baryon density 
$n_B/n_0\sim 6$ and dominates thereafter. At 
$T=200$ MeV the admixture of hadrons is, naturally, quite small.

Consider now penetrating probes. To find observable dilepton  
characteristics, one should integrate
the emission rate over the whole time-space $x\equiv (t,\bf x)$
evolution, add the contribution from the freeze-out surface
('hadron cocktail'), and take into account the experimental
acceptance. To simplify our task, we consider only the main
channel $\pi \pi \to \rho \to l^+ l^-$. In this case the dilepton 
emission rate is
\begin{eqnarray}
\frac{d^4 R^{l^+l^-}}{dq^4} =-\int d^4 x \ {\cal L}(M) \
\frac{\alpha^2}{\pi^3 q^2} \ f_B(q_0,T(x))   \
{\rm Im} \Pi_{em}(q,T(x),\mu_b(x))~, \label{rate1}
\end{eqnarray}
where  the integration is carried out over the whole space grid and time
from $t=0$ till the local freeze-out moment. Here
$q^2=M^2=q_0^2-{\mathbf q}^2$, $f_B(q_0, T(x))$ is the Bose distribution function,
and ${\cal L}(M)$ is the lepton kinematic factor.  The imaginary part of the
electro-magnetic current correlation function ${\rm Im} \Pi_{em}(q,T(x),\mu_b(x)$ 
includes in-medium effects which may be calculated in different scenarios.
The recent precise measurements of muon pairs~\cite{NA60} allowed one to discriminate 
two main scenarios, in particular those based on the Brown-Rho (BR) scaling
hypothesis~\cite{BR91} assuming a dropping $\rho$ mass and on a
strong broadening of $\rho$-meson spectral function as found in the many-body approach by Rapp and
Wambach~\cite{RW00}. It was shown~\cite{NA60} that the measured excess 
of muons is nicely described  by the strong broadening of the $\rho$-meson spectral 
function. In contrary, the BR scaling hypothesis predicts a large shift 
of the $\rho$-meson maximum towards lower invariant mass $M$ in contradiction 
with experimental data~\cite{NA60}.

\begin{figure}[h]
\centerline{\includegraphics[width=7cm]{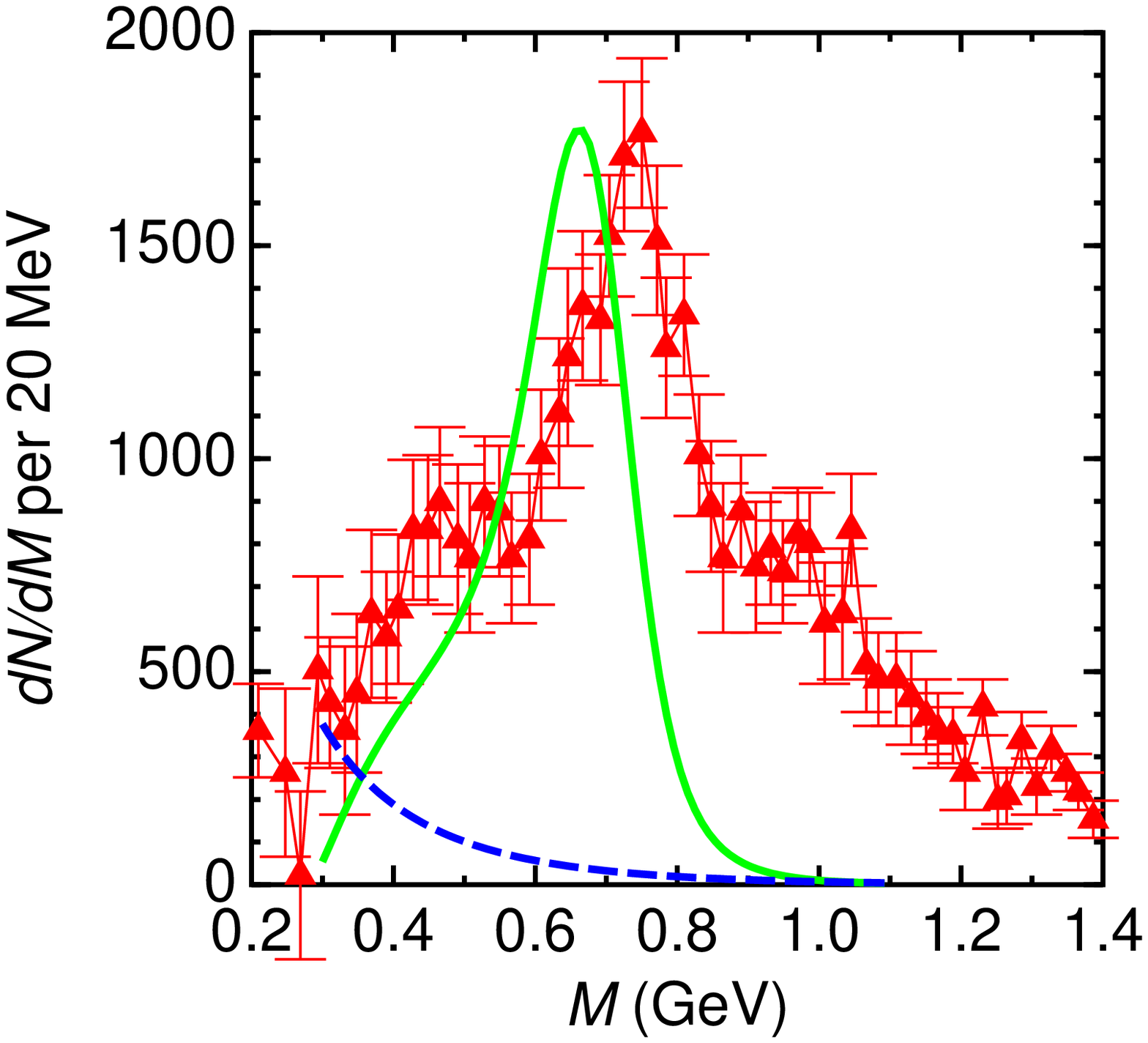}\hspace*{10mm}
\includegraphics[width=7cm]{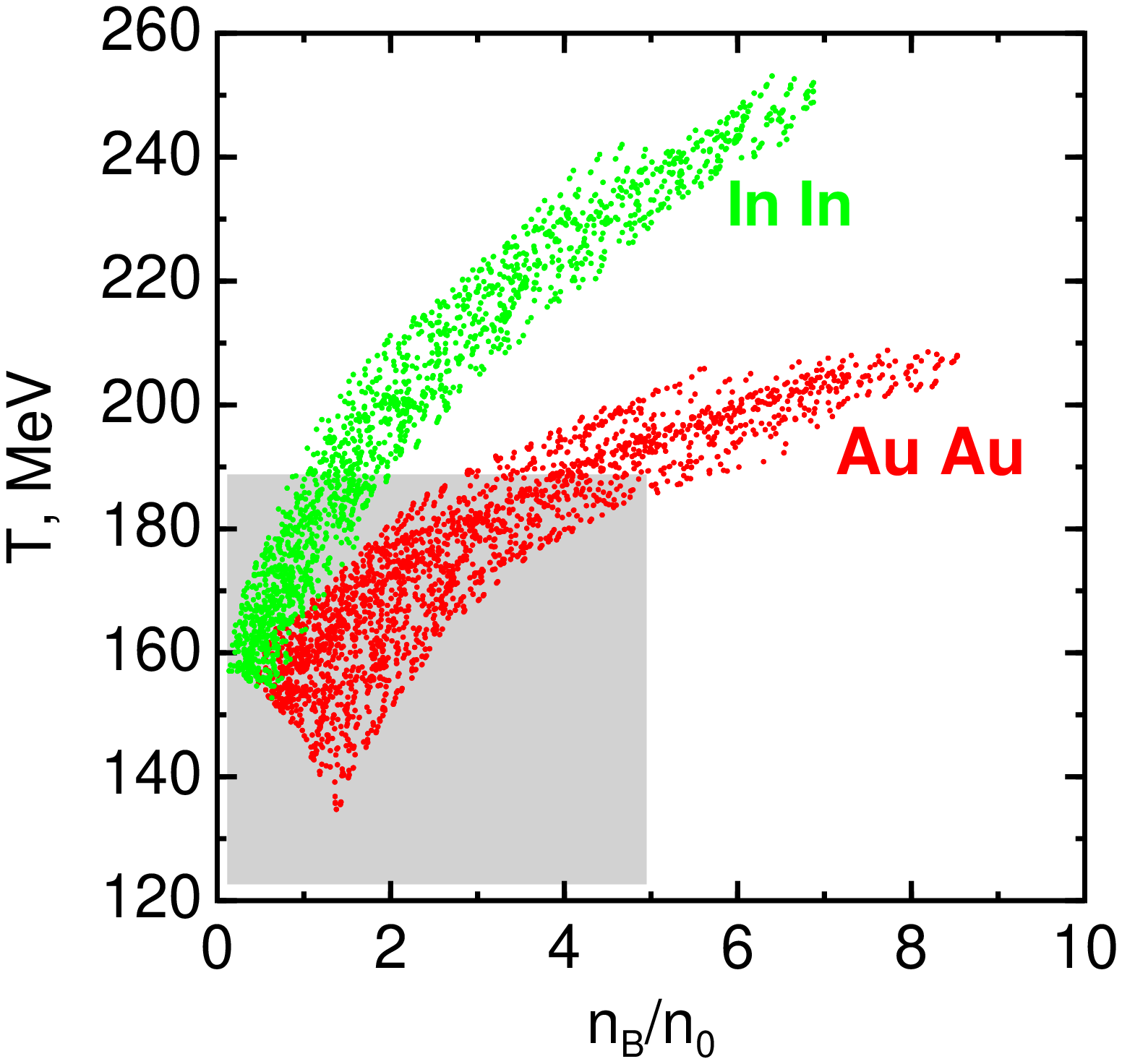}}
\caption{Invariant mass distribution of dimuons (left panel) from semi-central
  In+In collisions at the beam energy 158 $A$GeV. Experimental points are
  from~\cite{NA60}. The solid line corresponds to the $T$-independent dropping 
  mass~\cite{ST06}  and the dashed one is the contribution of the quark-hadron 
  bremsstrahlung channel. In the right panel one compares dynamical trajectories 
  projected onto the $T-n_B$ plane for central  In+In(158 AGeV) and Au+Au(40 AGeV). 
  The shaded region roughly corresponds to the hadronic phase.
 }
 \label{muon}
\end{figure}
This result looks quite disappointing. First, one sees no signal of a 
partial restoration of the chiral symmetry for the sake of which dilepton 
measurements were originally undertaken.
Second, to be consistent with the QCD sum rules {\it both} collision broadening 
{\it and} $\rho$-mass dropping should be taken into account~\cite{RRM06}. Generally, 
the exact relation between the $\rho$ mass and quark condensate is not fixed by 
the QCD sum rules in contrast with the BR scaling, so it is questionable whether 
to prescribe some $T$ dependence to the BR mass shift. This shortcomings of the 
analysis in~\cite{NA60} are commented by Brown and Rho~\cite{BR05}. 
In this respect, for the pion annihilation $\pi\pi \to \rho \to \mu^+\mu^-$ we 
estimated~\cite{ST06}  the imaginary part  of the 
$\rho$-meson self-energy in the one-loop approximation assuming the 
Hatsuda-Lee relation for the modified $\rho$ mass: 
$ m^*_{\rho}(x) =  m_{\rho} (1-0.15 \cdot n_B(x)/n_0)$.
The calculated result is presented in Fig.\ref{muon} by the 
solid line~\cite{ST06}. Indeed, the shift of the $\rho$-meson 
spectral function is not so drastic as in \cite{NA60}. 
In addition, one should note that dileptons carry direct information 
on the $\rho$ meson spectral function only
if the vector dominance is valid~\cite{BR05}. It is not the case in
the Harada-Yamawaki vector manifestation of hidden local symmetry~\cite{HY03}.
 
 As seen from Fig.\ref{muon}, the muon yield is underestimated at both low and high 
values of the $\mu^+\mu^-$ invariant mass $M$. In the  broadening scenario a low $M$ 
component is explained by particle-hole excitations but not a partial chiral 
symmetry restoration.  However, the existence of 
the mixed quark-hadron  phase, those study  is the main aim of the 
Nuclotron-based Ion Collider fAcility (NICA) 
project~\cite{NICA}, may give rise to a new dilepton source, quark(antiquark)-hadron 
bremsstrahlung.  Similarly to the $np$ bremsstrahlung, the process for an 
antiquark-hadron collision may   roughly be estimated in 
the soft-photon approximation as
\begin{equation}
\frac{dN_{qN}^{l^+l^-}}{dM^2}(s,M)\approx K \ \frac{\alpha^2}{3\pi^2} \ 
\frac{\bar{\sigma}(s)
} {M^2} \ \ln\left[\frac{s^{1/2}-m_N-m_q}{M}\right]~.
\label{qN}
\end{equation}
Here the averaged cross section $\bar{\sigma}(s)=\sigma_{el}^{qN}\ [s/(m_N+m_q)^2-1]$ 
and the elastic $qN$ cross section is approximated by the quark scaled $NN$ cross 
section~\cite{GK}
\begin{equation}
\sigma_{el}^{qN}=\left[\frac{18m_N (mb\cdot GeV^2)}{s-(m_N+m_q)^2}-10 \ (mb)\right]
\times \frac{1}{3}~.
\end{equation}
So the production rate will be
\begin{equation}
\frac{dN^{l^+l^-}_{qN}}{dM^2}=\int d^4x \int \frac{d^3 k_q}{(2\pi)^3} f({\bf 
k}_q,T(x)) \ \int \frac{d^3 N_q}{(2\pi)^3} f({\bf k}_N,T(x))  \ 
\frac{dN_{qN}^{l^+l^-}}{dM^2}
(s,M)\ v_{rel}~,
\label{rate}
\end{equation}
where $v_{rel}$ is the relative velocity of colliding $qN$ particles 
and the integration in 
(\ref{rate}) should be carried out over the whole space-time available for 
the mixed phase. The free mass is used for a nucleon and $m_q=150$ MeV for an 
antiquark. In a real case one should also add contributions from all other baryons, as 
well as that from quark-antibaryon and quark(antiquark)-meson interactions with proper 
cross sections. All these uncertainities are effectively introduced in (\ref{qN}) by
an arbitrary factor $K$.  

As follows from the dashed line in Fig.\ref{muon}, at rather reasonable value of
$K=10$ the $qN$ bremsstrahlung source improves agreement with experiment. It is of 
interest that the contribution of this new source decreases when the bombarding energy 
goes down till the NICA energy range ($\leq 40$ AGeV) while the contribution from 
particle-hole excitation is expected to grow since the baryon density in this range is 
higher (see below) allowing, in principle, disentangling of these two sources. 
The underestimated yield at high $M$, intermediate mass dileptons, can mainly be 
described by Drell-Yan process in the quark phase~\cite{Zah}. The mixed quark-hadron 
phase should also contribute to the intermediate $M$ region. 
Its contribution can be taken into account in the way as the Drell-Yan 
process with an additional hadron form factor. Effectively, it 
will increase the Drell-Yan lepton yield~\cite{Zah}. 

The phase distribution of all space cells at the early evolution moment $t=0.3$ fm/c 
projected on the $T-n_B$ plane is presented in Fig.\ref{muon}. It is seen that at the 
maximal NICA energy the baryon density in the hadronic phase for central Au+Au collisions 
is noticeably higher than that at the SPS energy in In+In collisions. It means that 
the difference between the BR scaling and broadening 
scenarios is expected to be more pronounced at the NICA energy.
\begin{figure}[h]
\centerline{\includegraphics[width=3.cm,angle=-90]{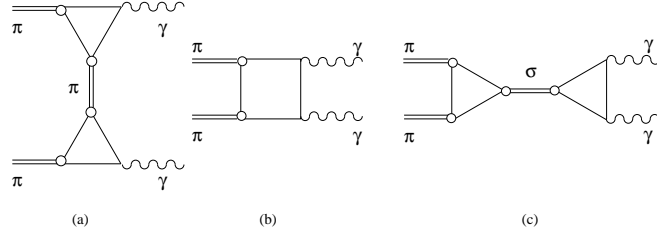}}
\caption{Quark diagrams for the 2$\gamma$ production  in the Born approximation
(a), (b) and through the $\sigma$ resonance (c).}
\label{diag}
\end{figure}

Annihilation of two pions into two photons is of particular interest since its cross 
section is sensitive to changes of the $\sigma$-meson properties which occur in the 
vicinity of chiral restoration phase transition. With increasing temperature and
density the $\sigma$ meson changes its character from a broad resonance with a large 
decay width into two pions to a bound state below the two-pion threshold
$m_\sigma (T,\mu_B) \lsim 2m_\pi(T,\mu_B)$. The calculation of 
the photon pair production rate at the given T as a function of 
the invariant mass shows a strong enhancement and narrowing of 
the $\sigma$ resonance at the threshold due to chiral symmetry 
restoration~\cite{VKBRS98}. We make the first estimate of this channel for 
a particular nuclear collision.

\begin{figure}[h]
\centerline{\includegraphics[width=6cm]{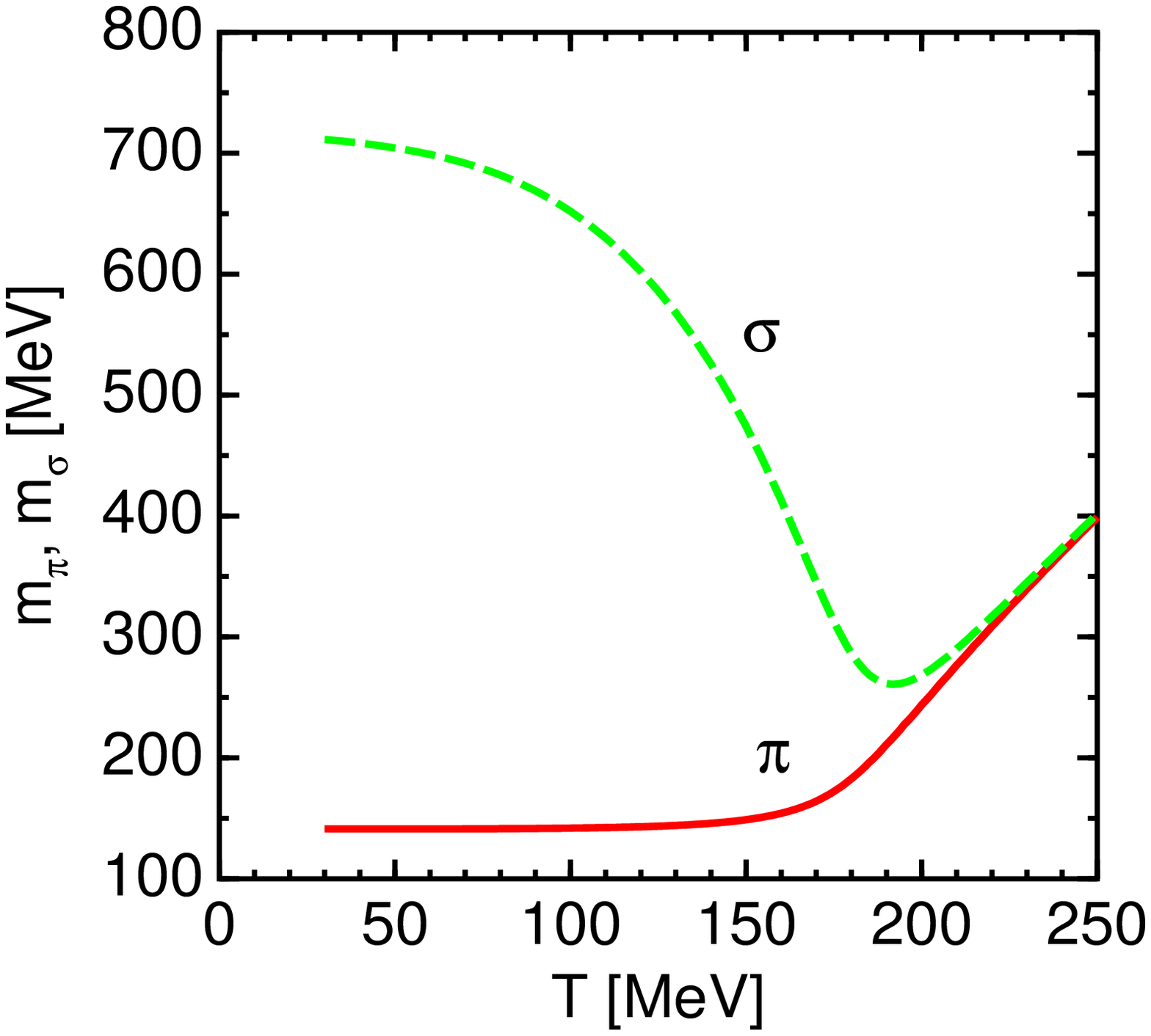}\hspace*{10mm}
\includegraphics[width=6cm]{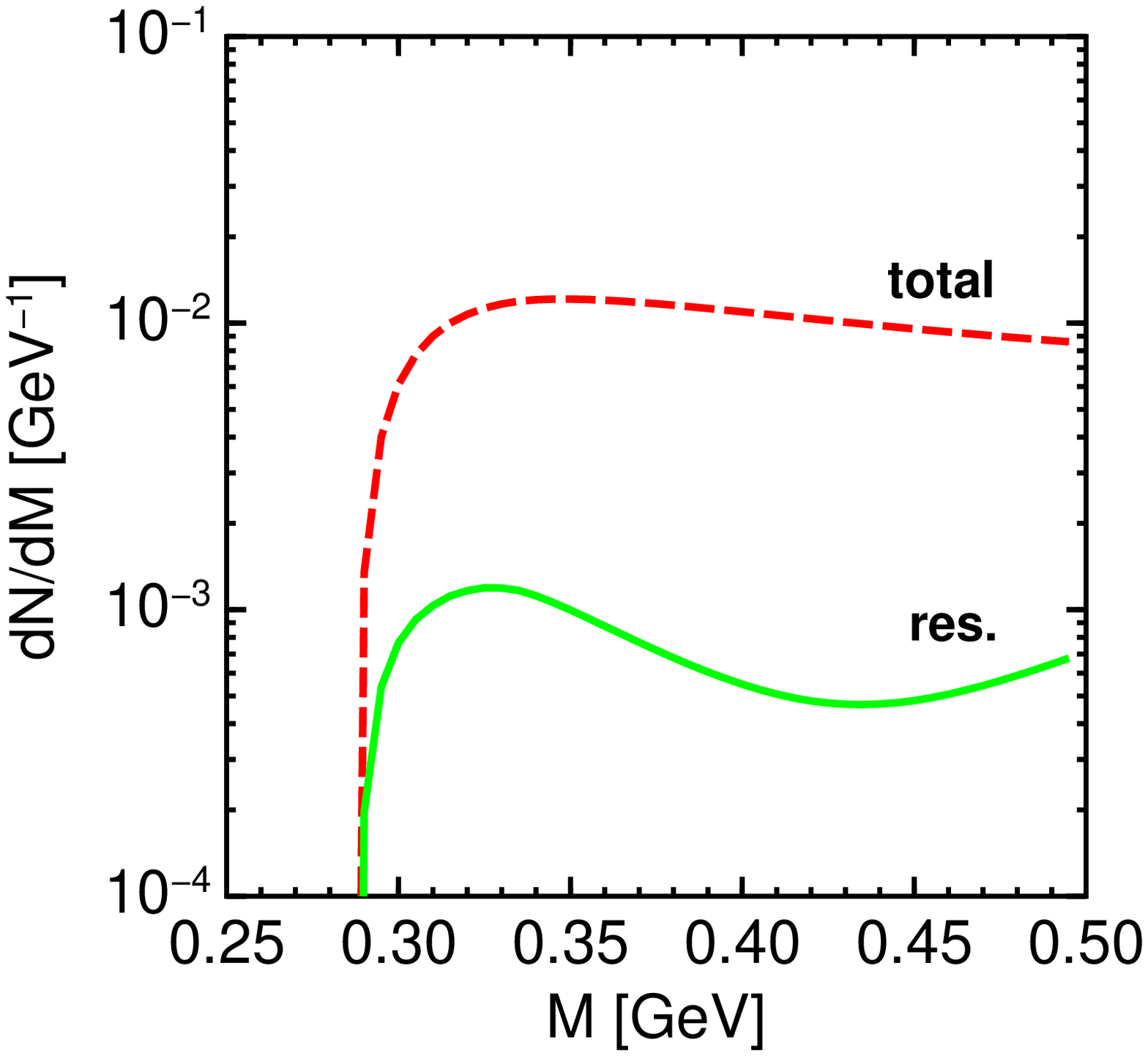}}
\caption{T-dependence of $\sigma$ and $\pi$ masses (left panel)
and total (dashed line) and resonance (solid line) invariant mass distributions 
of two photon pairs 
created in the central Au+Au collision at 40 AGeV (right panel).}
\label{phot}
\end{figure}
In-medium $\pi\pi \to \gamma\gamma$ process is evaluated within the NJL 
model~\cite{VKBRS98}. Besides the resonance diagram,  the dominating Born 
terms are considered, see Fig.\ref{diag}. So the total rate has the Born term, 
resonance term and interference 
 between them: $dN_{tot}^{\gamma\gamma}/dM^2=dN_{Born}^{\gamma\gamma}/dM^2
 +dN_{res}^{\gamma\gamma}/dM^2+dN_{interf}^{\gamma\gamma}/dM^2$
which should be substituted in eq.(\ref{rate}).

The temperature dependence of $\sigma$ and $\pi$ masses for this model is 
shown in Fig.\ref{phot}. The regime $m_\sigma (T,\mu_B) < 2m_\pi(T,\mu_B)$ starts
at $T\sim 165$ MeV. The photon yield as a function of photon pair invariant mass is 
presented in  Fig.\ref{phot} for central Au+Au collisions at 40 AGeV. The total number
of photon pairs sharply increases above the threshold $2m_\pi$ and then flattens on the 
level of $\sim 10^{-2}$. The resonance channel of interest is lower by about the order of 
magnitude as compared to the total yield  and exhibits a spread weak maximum. 
The maximum 
predicted in~\cite{VKBRS98} for the fixed T  is washed out, as seen from Fig.\ref{phot}.
It is not an easy but promising experimental problem to select out this maximum from 
the total distribution. Note that such an analysis should be carried out on a huge 
background of $\gamma$ decays of 'hadron cocktail'.

One of the authors (V.T.) is deeply indebted to D.I.~Blokhintsev who many years ago 
brought him into physics of high-energy interactions. We are thankful to  
E.~Kolomeitsev for
valuable remarks.    This work was supported in part by 
the Deutsche Forschungsgemeinschaft (DFG project 436 RUS 113/558/0-3),  the
Russian Foundation for Basic Research (RFBR grants 06-02-04001 and
08-02-01003), special program of the Ministry of Education and Science of the 
Russian Federation (grant RNP.2.1.1.5409).


\begin{thebibliography}{99}\itemsep -1mm
\bibitem{Bl57}D.I.~Blokhintsev, JETP {\bf 32}, 350 (1957).
%
\bibitem{ST06}V.V.~Skokov and V.D.~Toneev, Phys. Rev. {\bf C73}, 021902 (2006);
 Acta Physica Slovaca {\bf 56}, 503 (2006).
%
\bibitem{QGSM}
N.S.~Amelin, K.K.~Gudima, S.Y.~Sivoklokov and V.D.~Toneev,
  Sov. J. Nucl. Phys.  {\bf 52}, 172 (1990);
N.S.~Amelin, E.F.~Staubo, L.P.~Csernai, V.D.~Toneev and
K.K.~Gudima, Phys. Rev. C {\bf 44},  1541 (1991);
V.D.~Toneev, N.S.~Amelin, K.K.~Gudima and S.Yu.~Sivoklokov, Nucl. Phys.  
{\bf A519}, 463c (1990).
%
\bibitem{ST05} V.V.~Skokov and V.D.~Toneev, Sov. J. Nucl. Phys. {\bf 70}, 109 (2007).
%
\bibitem{TFNFR04} V.D.~Toneev, E.G.~Nikonov,  B.~Friman, W.~N\"orenberg,
and K.~Redlich, Eur. Phys. J. {\bf C32}, 399 (2004).
%
\bibitem{Fodor03} Z.~Fodor, Nucl. Phys. A {\bf 715}, 319 (2003);
 F. Csikor, G.I. Egri, Z. Fodor, S.D. Katz, K.K.
Szabo, and A.I. Toth, JHEP {\bf 405}, 46 (2004).
%
\bibitem{NA60}
NA60 Collaboration,	Phys. Rev. {\bf 96}, 162302 (2006); Nucl. Phys. {\bf A774}, 715 
(2006).
%
\bibitem{BR91}
G.E.~Brown and M.~Rho, Phys. Rev. Lett. {\bf 66}, 2720 (1991).
%
\bibitem{RW00}
R.~Rapp and J.~Wambach, Adv. Nucl. Phys. {\bf 25}, 1
(2000); R.~Rapp, G.~Chanfray and J.~Wambach,
  Phys.\ Rev.\ Lett.\  {\bf 76},  368  (1996)
%
\bibitem{RRM06} J.~Ruppert, T.~Renk and B.~Muller, Phys. Rev. {\bf C73}, 034903 
(2006). 
%
\bibitem{BR05}
G.E.~Brown and M.~Rho, nucl-th/0509001, nucl-th/0509002.
%
\bibitem{HY03} M.~Harada and K.~Yamawaki, Phys. Reps. {\bf 381}, 1 (2003).
%
\bibitem{NICA} NICA project: http://nica.jinr.ru.
%
\bibitem{GK} C.~Gale and J.~Kapusta, 
Phys. Rev. {\bf C35}, 2107 (1987).
%
\bibitem{Zah} C.M.~Hung and E.V.~Shuryak, K.~Dusling,  Phys. Rev. {\bf C56}, 453 (1997); 
D.~Teaney and I.~Zahed, Phys. Rev. {\bf C75}, 024908 (2007); H. van Hees and Rapp, arXiv:0711.3444.
%
\bibitem{VKBRS98}M.K.~Volkov, E.A.~Kuraev, D.~Blaschke, G.~R\"{o}pke and S.~Schmidt,
Phys. Lett. {\bf B424}, 235 (1998).
%
\end{thebibliography}
\end{document}